\documentclass{ws-procs975x65}

\usepackage{color}
\newcommand{\MPl}{M_{_{\rm P}}}
\def\beq{\begin{equation}}
\def\eeq{\end{equation}}
\def\bea{\begin{eqnarray}}
\def\eea{\end{eqnarray}}
\def\benu{\begin{enumerate}}
\def\eenu{\end{enumerate}}
\def\nn{\nonumber}

\def\l{\left}
\def\r{\right}
\def\lp{L_{_{\rm P}}}

\begin{document}

\title{Quantum gravitational corrections to the propagator\\ 
in spacetimes with constant curvature}
\author{Dawood Kothawala$^{(1)}$, 
S.~Shankaranarayanan$^{(2,3)}$\footnote{Speaker,
E-mail:~shanki@iisertvm.ac.in}, and L.~Sriramkumar$^{(4)}$}
\address{${}^{(1)}$ IUCAA, Post Bag 4, Ganeshkhind, Pune 411 007, India.\\
${}^{(2)}$~Institute of Cosmology and Gravitation, University of Portsmouth, Portsmouth, U.K.\\
${}^{(3)}$ School of Physics, Indian Institute of Science Education and Research-Trivandrum,\\
CET campus, Thiruvananthapuram 695 016, India.\\
${}^{(4)}$ Harish-Chandra Research Institute, Chhatnag Road, Jhunsi, 
Allahabad 211 019, India.}

\begin{abstract}
  The existence of a minimal and fundamental length scale, say, the
  Planck length, is a characteristic feature of almost all the models
  of quantum gravity.  The presence of the fundamental length is
  expected to lead to an improved ultra-violet behavior of the
  semi-classical propagators.  The hypothesis of path integral duality
  provides a prescription to evaluate the modified propagator of a
  free, quantum scalar field in a given spacetime, taking into account
  the existence of the fundamental length in a locally Lorentz
  invariant manner.  We use this prescription to compute the quantum
  gravitational modifications to the propagators in spacetimes with
  constant curvature, and show that: (i)~the modified propagators are
  ultra-violet finite, and (ii)~the modifications are non-perturbative
  in the Planck length.  We discuss the implications of our results.
\end{abstract}


\bodymatter

\section{Motivation}

It is presumed that quantum gravitational effects would become
important at length scales of the order of the Planck length, $\lp =
(G\, \hbar /c^3)^{1/2}$.  At these scales, it seems quite likely that
the description of the spacetime structure in terms of a metric, as
well as certain notions of standard quantum field theory, would have
to undergo drastic changes.  Since any quantum field has virtual
excitations of arbitrary high energy, which probe arbitrarily small
scales, it follows that the conventional quantum field theory can only
be an approximate description that is valid at energies smaller than
the Planck energy. In particular, one hopes that, in a complete
theory, gravity would provide an effective cut off at the Planck
scale. Some very general considerations based on the principle of
equivalence and the uncertainty principle seem to strongly indicate
that it may not be possible to operationally define spacetime events
beyond an accuracy of the order of $\lp$\cite{1987-Padmanabhan-CQG}.
Therefore, one may consider $\lp$ as the `zero point length' of
spacetime intervals.

The existence of a fundamental length implies that processes involving
energies higher than the Planck energies will be suppressed, thereby
improving the ultra-violet behavior of the theory.  However, according
to a theorem due to Weinberg, the momentum space propagator of any
Lorentz invariant and local field theory {\it has}\/ to behave
as~$p^{-2}$ in the ultra-violet limit~\cite{1979-Weinberg-Proc}.
Therefore, if the short-distance behavior of the propagators have to
be improved, one has to either break Lorentz invariance or include
non-local terms in the field theory.  In this talk, we shall focus on
a latter approach that is based on the {\it hypothesis of
  path-integral duality} \cite{1997-Padmanabhan-PRL}.  In flat
space-time, it has been shown that the modified propagators obtained
using the duality principle are Lorentz invariant and ultra-violet
finite.  We shall extend this analysis to space-times with constant
curvature and explicitly show that the two-point function of the
scalar field in these space-times are finite in the coincident
limit~\cite{2009-Kothawala.etal-PRD}.

\section{The hypothesis of path-integral duality}

The basic postulate is that the path integral amplitude of a point
relativistic particle is invariant under the `duality transformation'
${\cal R} \rightarrow (\lp^2/ {\cal R})$, where $\mathcal{R}$ is the
proper length of the path~\cite{1997-Padmanabhan-PRL}.  The specific
prescription being that the original sum over the paths is modified to
\begin{equation}
G_{_{\rm PID}}(x,x') 
=\sum_{\mathrm{paths}} 
\exp{-m \left({\cal R}(x,x')
+ \frac{{\lp^2}}{{\cal R}(x,x')}\right)}, 
\end{equation}
where $m$ is the mass of the relativistic particle.
%
It can be shown that the resulting, modified momentum space propagator 
has the following forms in the infra-red and the ultra-violet limits:
\begin{eqnarray}
G_{_{\rm PID}}(p)\propto \l\{\begin{array}{ll}
(p^{2} + m^2)^{-1},\;\;\;\; &{\rm for }\;\;\;\; 
(\lp\, \sqrt{p^2 + m^2}) \ll 1,\\
\frac{\exp-\l(\lp\, \sqrt{p^2 + m^2}\r)}{\lp^{1/2}\, (p^2 + m^2)^{3/4}},
\;\;\;\; & {\rm for }\;\;\;\; (\lp\, \sqrt{p^2 + m^2}) \gg 1.\\
\end{array}\r.
\end{eqnarray}
It is interesting to note that the modified propagator probably 
corresponds to a field theory with infinite derivatives and is 
therefore likely to be highly non-local~\cite{1999-Padmanabhan-MPLA}. 


\section{Modified propagator in constant curvature space-times}

As we mentioned, in flat space-time, the hypothesis of path-integral
duality renders the propagator ultra-violet finite.  It is then
natural to enquire whether the hypothesis performs in a similar
fashion in an arbitrary curved space-time as well.  It turns out to be
a formidable tasks to compute the propagator in an arbitrary
background.  To make the calculations tractable, we focus on
space-times of constant curvature,
$R\propto\ell^{-2}$~\cite{2009-Kothawala.etal-PRD}.  In particular, we
consider (i)~the Einstein static spacetime in $(3+1)$-dimensions [i.e.
${\rm R} \times {\rm S}^3$], (ii)~the de Sitter and the anti-de Sitter
spacetimes in $(3+1)$-dimensions [i.e.  Euclidean ${\rm S}^4$ and
${\rm H}^4$], and (iii)~the anti-de Sitter spacetime in
$(2+1)$-dimensions [i.e.  Euclidean ${\rm H}^3$].

Upon using the Schwinger's proper time representation, we find that 
we can evaluate the modified propagator for a massive scalar field 
in the above backgrounds (for details, see 
Ref.~\cite{2009-Kothawala.etal-PRD}). 
The main results can be summarized as follows $\left[b = \left(m^2 
+ (\xi - \frac{1}{6})\, R \right);  
\beta = (1 + (m\, \ell)^2 + \xi\, R\, \ell^2) \right]$:
\begin{enumerate}
\item In $(3+1)$-dimensions:
{\small 
\beq
\!\!\!\!\!\!\!\!\!\! 
G_{_{\rm PID}}(x, x') 
= -\l(\frac{\sqrt{b}}{8 \pi}\r)\; 
\frac{H_{1}^{(2)}\l(\sqrt{b\, 
\l[u_{xx'}^2 -\lp^2\r]}\,\r)}{\sqrt{u_{xx'}^2
- \lp^2}}
\; \Delta_{xx'}^{1/2} 
\underset{x \rightarrow x'}{\longrightarrow}
-
\frac{\sqrt{b}}{8\, \pi\, i\, \lp}
H_1^{(2)} \l(i\, \sqrt{b}\, \lp\r),\nn
\eeq
}
\item
In $(2+1)$-dimensions:
{\small 
\beq
\!\!\!\!\!\!\!\!\!\! 
G_{_{\rm PID}}(x,x') 
= \l(\frac{1}{4\, \pi\,}\r)\;  
\frac{\exp{-\sqrt{ (\beta / \ell^2) \, \l[u_{xx'}^2 + \lp^2\r]}}}{\sqrt{u_{xx'}^2
+ \lp^2}}
\Delta_{xx'}^{1/2} 
\underset{x \rightarrow x'}{\longrightarrow} 
\frac{1}{4\, \pi\, \lp}
\, \exp{-\l(\sqrt{\beta}\; \frac{\lp}{\ell} \r)},\nn
\eeq
}
\end{enumerate}
where $u_{xx'}$ represents the geodesic distance between the two
points $x$ and $x'$, and $\Delta_{xx'}$ is the Van-Vleck determinant.
We see that the duality hypothesis: (i) regulates the theory at Planck
scales, (ii) yields modifications which are non-perturbative in~$\lp$
and, (iii) most interestingly, the quantum gravitational effects, as
accounted for by the duality prescription, can be looked upon as
leading to addition of $\lp$ to {\it all}\/ spacetime (geodesic)
intervals in a (peculiar) Pythagorean way, i. e. , $\l\langle
\sigma^2(x,x')\r\rangle = [\sigma^2(x,x') + {\cal O}(1)\, \lp^2]$, as
is evident from the above expressions for the
propagators. (here, the angular brackets represent a suitable
path integral average over quantum fluctuations of the background metric.)

\section{Implications}

Current approaches to quantum gravity seem to face two main obstacles:
(i)~construct a consistent description of physics at Planck scales,
and (ii)~make robust predictions that can be tested against
experiments.  The hypothesis of path-integral duality accepts our
ignorance of physics at the Planck scale and instead, based on an
underlying principle, provides a prescription for calculating
gravitationally smeared propagators which can be used to make testable
predictions.  For instance, path-integral duality predicts that the
Planck scale corrections to the primordial perturbation spectrum in
exponential inflation will be of the order
of~$(H/\MPl)$\cite{2006-Sriramkumar.Shankaranarayanan-JHEP}.

Our results here seem to support the viewpoint that demanding the
duality invariance of the relativistic point particle path integral is
{\it equivalent}\/ to `adding' a zero-point length to spacetime
intervals.  Such a result might be an outcome of {\it the generic
  short distance behavior}\/ of the spacetime structure itself and
hence could be expected to naturally appear in the (effective) low
energy sector of the full theory of quantum gravity
\cite{2006-Fontanini.etal-PLB}.

\end{document}